# Spin Qubit Properties of the Boron-Vacancy/Carbon Defect in the Two-Dimensional Hexagonal Boron Nitride

*Sergey Stolbov[1] and Marisol Alcántara Ortigoza[2]

[1] *Physics Department, University of Central Florida, Orlando, Florida 32816, USA*

[2] *Physics Department, Tuskegee University, Tuskegee Institute, Alabama, 36088 USA*

*Corresponding author: Sergey.Stolbov@ucf.edu

**Abstract.** Spin qubit defects in two-dimensional materials have a number of advantages over those in three-dimensional hosts including simpler technologies for the defect creation and control, as well as qubit accessibility. In this work, we select the $V_BC_B$ defect in the hexagonal boron nitride (hBN) as a possible optically controllable spin qubit and explain its triplet ground state and neutrality. In this defect a boron vacancy is combined with a carbon dopant substituting the closest boron atom to the vacancy. Our density-functional-theory calculations confirmed that the system has dynamically stable spin triplet and singlet ground states. As revealed from our linear response GW calculations, the spin-sensitive electronic states are localized around the three undercoordinated N atoms and make local peaks in the density of electronic states within the bandgap. Using the triplet and singlet ground state energies, as well as the energies of the optically excited states, obtained from solution to the Bethe-Salpeter equation, we construct the spin-polarization cycle, which is found to be favorable for the spin qubit initialization. The calculated zero-field splitting parameters ensure that the splitting energy between the spin projections in the triplet ground state is comparable to that of the known spin qubits. We thus propose the $V_BC_B$ defect in hBN as a promising spin qubit.

# I. INTRODUCTION

Spin-active local defects in wide-bandgap semiconductors have become a subject of increasingly extensive studies because they are promising building blocks for quantum technology applications. More specifically, in the optically addressable defects, their spin states can be initialized, manipulated, and readout by using optical excitation and emission as well as microwave means. Such defects can serve as spin qubits for quantum computing and sensing. The most known spin-qubit defect is the NV$^-$ negatively charged center in diamond, which combines the nitrogen atom substitution of a carbon next to a carbon vacancy. [1, 2] This defect has a spin triplet ground state and a higher energy local-minimum singlet state. Due to the dipolar spin-spin interaction the triplet spin projections m$_S$ = 0 and m$_S$ = ±1 split (zero-field splitting) within the microwave range energy. If the triplet is optically excited, it undergoes the phonon-assisted intersystem crossing (ISC) transition to an excited state of the singlet configuration. This specific ISC obeys a selection rule in which only triplet states with m$_S$ = ±1 can transition from the triplet to singlet state. The latter process is followed by the optical emission from the singlet excited to its ground state and the nonradiative transition back to the triplet ground state. Simultaneously, the optical emission from the remaining triplet m$_S$ = 0 excited state to its ground state occurs. This spin-polarization cycle thus results in the qubit initialization of the triplet at m$_S$ = 0 state, [2] which is ready for further optical and microwave manipulation and readout.

The other well-known spin-qubit defects are the negatively charged Si vacancy and the Si – C divacancy in SiC. [3, 4] Although the total spin of the Si vacancy is 3/2, the spin-polarization cycle in these defects is similar to that in the NV$^-$ center. In a recent computational work, [5] the authors proposed a new defect with qubit functionality. That is the aluminum-vacancy/sulfur complex (V$_{Al}$S$_N$) in the wurtzite AlN. The first-principles evaluation of the electronic structure and optical excitations in V$_{Al}$S$_N$ indicates that the defect is capable to undergo the spin-polarization cycle similar to that in the NV$^-$ center.

The above defects are hosted in three-dimensional (3D) semiconductors. Meanwhile, since recently, the search for defect with the spin-qubit functionality has been extended to two-dimensional (2D) systems with a main focus on defects in hexagonal boron nitride (hBN). This is a wide-bandgap (6 eV) layered semiconductor with easily exfoliated monolayers. Hexagonal BN is known, in particular, as a host for single-photon emitters. [6] 2D systems have various advantages over 3D structures. It is easy to create and control the desired defect in 2D hosts and to integrate them into a setup. As for quantum sensing applications, the proximity of a 2D sensor to a sample can significantly increase its sensitivity. [7] The first identified spin defect in hBN with the properties favorable for spin qubit functionality was the negatively charged boron vacancy ($V_B^-$). [8] The authors found that $V_B^-$ has a triplet ground state and demonstrated a spin polarization cycle and optical readout. This work has been followed by several experimental and computational studies of this defect [9 – 11], which revealed that, similar to the Si vacancy in SiC, the polarization cycle does not include and does not require optical emission to the singlet ground state. The defect has a low spin coherence time which results from the interaction of the defect states with the nuclear spin bath of the host atoms. The disadvantage of this spin qubit is its very low quantum yield. [9] Other defects in hBN have been also under consideration. The authors of a computational work proposed the carbon tetramer defect in hBN as possible spin qubit. [12] According to their computational results, the tetramer defect has both triplet and singlet states and is capable to undergo a spin polarization cycle. Another recent computational and experimental study [13] reveals that the B asymmetric divacancy also has the triplet and singlet states. However, in this system, the triplet and singlet states are almost degenerate which makes the ISC from the singlet to triplet state forbidden.

In this work we put to the test a $V_BC_B$ defect in hBN. This defect has been reported to be a triplet in the database of Ref. 14. However, neither the specific structure, nor the rationale for its triplet ground state, nor its electronic structure, nor the spin-qubit functionality or lack of it has been addressed so far. By applying the existing knowledge, we focus on explaining why the complex combining a boron vacancy and another boron atom substituted by carbon ($V_BC_B$) has a *neutral* triplet ground state, and demonstrate that its properties are suitable for the spin qubit functionality. We perform density-functional-theory (DFT)-based calculations to find the triplet and singlet states and test their dynamical stability. Next, we apply the linear response GW [15] and Bethe-Salpeter equation (BSE) [16] methods to evaluate the defect's electronic structure and optical excitations followed by the construction of the spin-polarization cycle and the calculation of the zero-field splitting.

## II.  COMPUTATIONAL DETAILS

To perform the first-principle calculations we used the Vienna Ab-initio Simulation Package VASP 6.4 [17] with the projector-augmented-wave potentials. [18] The 400-eV cutoff energy for the plane-wave expansion was used for all calculations. The total energy of the systems and phonon spectra (at the Γ-point) were calculated within DFT with the Perdew-Burke-Ernzerhof (PBE) approximation for the exchange-correlation functional. [19] Periodicity of the defected system was achieved by using the 5x5x1 50-atom supercell. The Brillouin zone was sampled with the 3x3x1 k-point mesh. The electronic structure, in terms of the independent quasi-particle (IQP) states, was calculated using the GW method [15] within the $G_0W_0$ approximation. The obtained GW wavefunctions and kernels were used to calculate the frequency-dependent dielectric functions and oscillator strengths of the optical excitations within BSE. [16]

## III.  RESULTS AND DISCUSSION

The first requirement for a spin qubit is that the corresponding defect must have a spin-polarized ground state (S=1 or higher). It is known that, in solids, weak chemical bonds or, better, broken bonds between an atom and its neighbors are favorable for formation of non-zero spin state of the atom. The easiest way to break the bonds and make the neighboring atoms undercoordinated is the creation of a vacancy in crystal. For example, the creation of a B-vacancy in hBN makes three N atoms undercoordinated. And as a result, this defect becomes spin-polarized. [8] However, removal of the B atom reduces the number of valence electrons in the system by one (note that we consider only valence *p*-electrons and neglect the effects of semi-core 2s- electrons of B). That would make the number of valence electrons odd leading to a total spin equal to 3/2. If the defect became negatively charged (that is, one extra electron is added to the system) the number of electrons would be even resulting in a 2/2 total spin (triplet). This is the mechanism of formation of a triplet state in the $V_B^-$ defect in hBN. However, once it is understood why boron vacancies occur in a negatively charged state, one can devise another way to bring back the missing valence *p*-electron, while *keeping the defect electrically neutral*, which is to substitute another host atom with a dopant having a suitable number of the *p*-valence electrons. [5] Namely, in the case of the $V_B C_B$ defect, the substitution of another B atom with a carbon atom - which has two valence *p*-electrons – leaves three N atoms undercoordinated and keeps the number of valence *p*-electrons even while leaving the defect electrically neutral. In this way we explain why the $V_B C_B$ defect has a triplet ground state and is expected to be neutral. The above reasoning to find defects with triplet ground states has a caveat in regard to suitable substitutional dopants. Namely, if the radius of the dopant is significantly larger than that of the host atoms, it is likely that the ground state will be a singlet because the dopant will tend to compress the bonds among the host atoms and prevent the undercoordinated ones to become spin polarized.

### A. Spin states and stability of the $V_B C_B$ defect

The next steps are to perform the first-principles calculations to a) reveal the equilibrium geometric structure of the defect, b) check if the spin triplet state is the ground state, c) find whether the defect has also a singlet state, and d) evaluate stability of the defect. All these steps involve only electronic ground state processes; therefore, we perform the corresponding calculations within DFT.

One experimental method for the creation of carbon-doping/vacancy defects in hBN is the carbon ion implantation. [20] In this process, the impact of a ~10 keV-energy C ion is assumed to cause a B-vacancy accompanied by the substitution of another B atom with the incoming C ion. In light of the above process, we create a supercell with the $V_B C_B$ defect configuration shown in Fig. 1 and calculate the relaxed defect structure. Indeed, we confirmed that this defect has a triplet ground state. Considering that the C-dopant may take some other positions, we also consider another configuration with carbon substituting the B atom marked in the figure by white cross. We find, however, that the total energy of this configuration is 0.727 eV higher than that of the first configuration, which may be indicative of the role of carbon in a vacancy healing effect. Another option may be that the carbon atom substitutes, not B, but N, creating the $V_B C_N$ defect instead. However, it has been shown earlier that this defect is dynamically unstable and undergoes a transition to the stable $V_NC_B$ defect. [21] In Table 1, we thus compare the energetic characteristics of both neutral $V_BC_B$ and $V_NC_B$ defects. We provide the standard formation energy of the defects obtained as:

$$E_{form}(V_BC_B) = E_{tot}(V_BC_B) + 2\mu(B) - E_{tot}(BN) - \mu(C) \qquad (1)$$
$$E_{form}(V_NC_B) = E_{tot}(V_NC_B) + \mu(B) + \mu(N) - E_{tot}(BN) - \mu(C) \qquad (2)$$

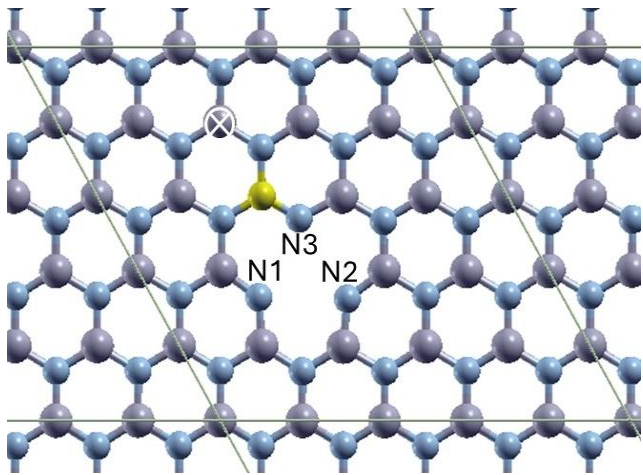

FIG. 1. The proposed configuration of the $V_B C_B$ defect. The grey, blue, and yellow balls represent the B, N, and C atoms respectively. N1, N2, and N3 mark three undercoordinated N atoms. The white cross marks the other considered position of the C atom. The bright green lines show the supercell boundaries.

Here, $E_{tot}(BN)$, $E_{tot}(V_BC_B)$ and $E_{form}(V_NC_B)$ are the total energy of the pristine and $V_BC_B$- and $V_NC_B$ defected hBN structures, respectively; and $\mu(\square)$ is the chemical potential of X (for X=B,N, and C).

Note in Table 1 that for $V_NC_B$, its formation energies of under N-poor and N-rich conditions are identical. This is expected because the only terms that vary in Eq.(2), between the N-poor and N-rich conditions, are the chemical potentials $\mu(B)$ and $\mu(N)$. Nevertheless, their sum $\mu(B) + \mu(N)$ remains constant and is equal to the chemical potential of the pristine hBN, $\mu(BN)$, under both conditions.

The formation energy is a thermodynamic quantity giving the energy of configurations achieved in infinite time. However, the experimental defect creation involves not only thermodynamic process but also, and more importantly, a kinetic process. Namely, the lifetime of a thermodynamically unstable defect is a kinetic quantity that depends on the height of the activation energy barrier(s) to reach the most stable configuration, which is determined by the strength of the chemical bonds between the dopant and its neighbors. Therefore, in addition to the formation energy, we provide in Table 1 the dopant binding energy ($E_B$) of the defect. For $V_BC_B$ it is

$$E_B(C) = E_{tot}(V_BC_B) - E_{tot}(2V_B) - E_{at}(C) \qquad (3)$$

And for $V_NC_B$

$$E_B(C) = E_{tot}(V_NC_B) - E_{tot}(V_BV_N) - E_{at}(C) \qquad (4)$$

Here, $E_{tot}(2V_B)$ is the total energy of defected structure with two B-vacancies and $E_{tot}(V_BV_N)$ is the total energy with of defected structure with one B-vacancy and one N-vacancy,

Table 1. Formation energy and binding energy of C calculated for the considered defects

| Defect | $E_{form}$ N-rich (eV) | $E_{form}$ N-poor (eV) | $E_B(C)$ (eV) |
|--------|------------------------|------------------------|---------------|
| $V_BC_B$ | +6.577 | +11.793 | -16.337 |
| $V_NC_B$ | +9.223 | +9.223 | -12.247 |

The absolute value of the C binding energy in $V_BC_B$ is much higher than that in $V_NC_B$, indicating, as expected, that breaking C-N bonds requires more energy that breaking C-B bonds, and hence suggesting that the lifetime of $V_BC_B$ has to be longer than that of $V_NC_B$. We thus conclude that the $V_BC_B$ defect is energetically more favorable than $V_NC_B$, and its preferred structure is the one shown in Fig. 1. Note that, due to the presence of the C dopant, the defect has an asymmetric structure in which, in contrast to the $V_B$ defect [8], the N1, N2, and N3 atoms are nonequivalent. To test the dynamic stability of $V_BC_B$, we calculate its phonon spectrum. We find that there is no vibrational mode with imaginary frequency, which indicates that the defect is dynamically stable.

To find a singlet state of $V_BC_B$ we start with the relaxed triplet structure and re-relax it without spin polarization. Then, we used the obtained structure as the input to rerun the structural relaxation with the spin polarization on. As a result, we obtained the system in the singlet states with total energy 0.043 eV higher than that of the triplet state. The phonon spectrum of the singlet confirms that it is also dynamically stable. To test the accuracy of the results we calculated both triplet and singlet DFT total energy for the smaller 4x4 supercell. We found that for the 4x4 case the difference between total energy of the triplet and singlet is 0.043506 eV, while for the 5x5 supercell it is 0.04315 eV. Moreover, the DFT total density of state of the 6x6 supercell is almost identical to that of the 5x5 supercell. We thus conclude that the 0.043 eV obtained for the 5x5 supercell is a reliable number for the energy difference.

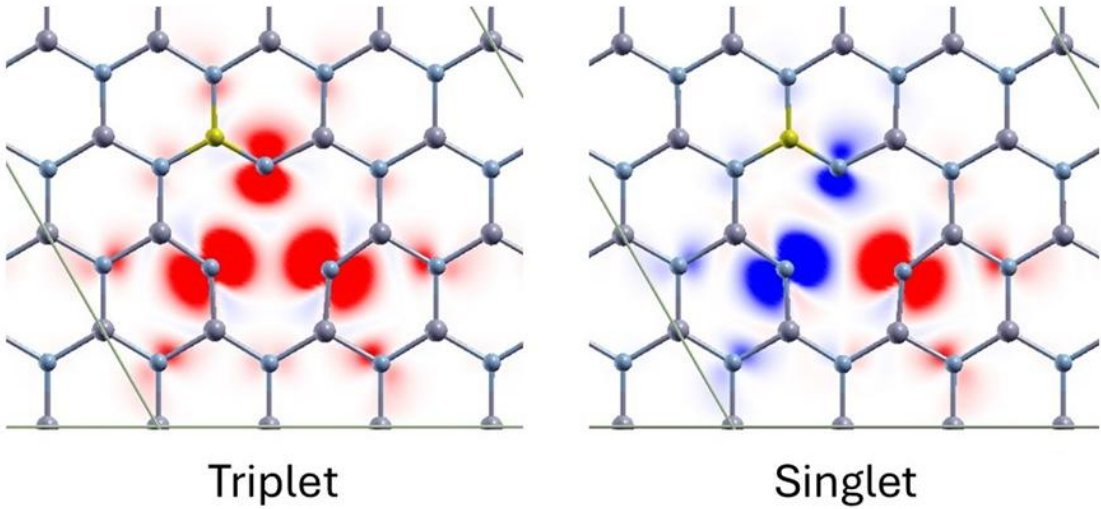

FIG. 2. The in-plane cut of the spin density calculated for the triplet and singlet states of the $V_BC_B$ defect. The red and blue areas correspond to the spin-up and spin-down densities, respectively.

Fig. 2 illustrates the spin density distribution in the $V_BC_B$ defect. As expected, the spin density is mostly localized around the undercoordinated N atoms. Interestingly, although the total spin of the singlet state is zero, the three undercoordinated N atoms are spin-polarized with the spin-up and spin-down contributions compensating each other. The asymmetry of the spin density reflects the defect structural asymmetry.

## B. Electronic structure and optical excitations in the $V_BC_B$ defect

We found that the $V_BC_B$ defect has a triplet ground state, as well as a singlet state with higher total energy. This is a necessary but not sufficient condition for the spin qubit functionality. The spin-polarization cycle is a process involving the excited states of the system, which we evaluate next. Since DFT fails to account for the excited states, we move at this point to the advanced GW and BSE methods which are sufficient to describe the properties in question.

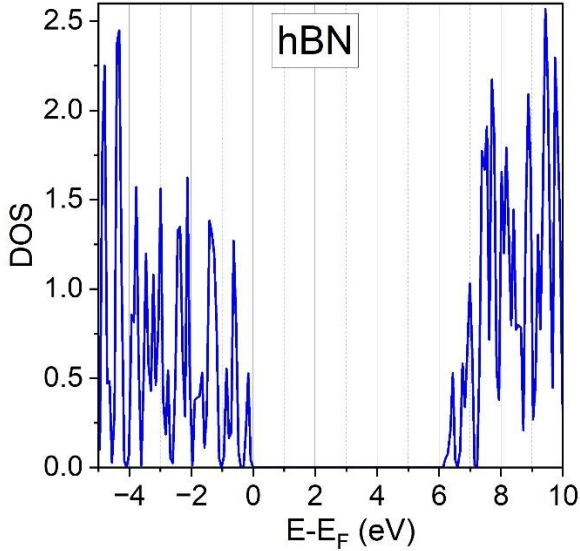

FIG. 3. Density of IQP states calculated for the pristine hBN primitive unit cell.

To test the accuracy of our GW settings and have a system of reference for the defect, we calculate the electronic structure of the pristine hBN. Within GW, the electronic structure is represented by the eigen states of independent quasiparticles (IQP) which are in fact electrons or holes screened by electron-electron interaction. Shown in Fig. 3 is the density of IQP states obtained for hBN from our GW calculations. One can see that the calculated bandgap width is ~6 eV, which is in good agreement with the experimental hBN gap.

Next, the GW calculations were performed for the triplet and singlet states of the $V_BC_B$ defect. As seen from Fig. 4, in contrast to the pristine hBN (Fig. 3), extra peaks related to the $V_BC_B$ defect appear in the density of states (DOS). The occupied peaks are located just above the valence band (VB) within ~1 eV below the Fermi level while the unoccupied ones are positioned within the bandgap, approximately 4 eV above the Fermi energy. As seen from Fig. 2, the spin polarized states for both triplet and singlet are formed around the undercoordinated N atoms: N1, N2, and N3. From the shape of the spin-density lobes, we conclude that they are formed upon an in-plane hybridization of the *p*-electron states of these atoms. To reveal their contribution to the total DOS of the system we analyze the local densities of the *p*-states of the N1, N2, and N3 atoms. As shown in Fig. 5, the occupied *p*-states of the N1 and N2 atoms form a wide low-density band reflecting their hybridization with the VB states, as well as narrow peaks corresponding to relatively isolated (not hybridizing) states just above VB, which are mostly responsible for the spin polarization in $V_BC_B$. The *p*-states of N3 exhibit an enhanced peak between 1 and 3 eV below the Fermi level, which corresponds to a strong hybridization with the neighboring C atom causing the reduced spin polarization at the N3 atom. In the singlet case, the *p*-states of the undercoordinated N atoms form extra sharp occupied DOS peaks at 2 eV below the Fermi level in addition to the states between -1 and 0 eV. We conclude that, even though the peaks at -2 eV overlap with host's VB, their sharpness tells us that there is no significant hybridization. Consequently, the peaks at -2 eV contribute to the local spin polarization of the singlet. The narrow unoccupied DOS peaks are also associated with the *p*-states of the N1, N2, and N3 atoms for both triplet and singlet. Such an electronic structure seems to be favorable for sharp-peak optical excitation and emission. Thus, the next step is focused on the optical properties of the $V_BC_B$ defect.

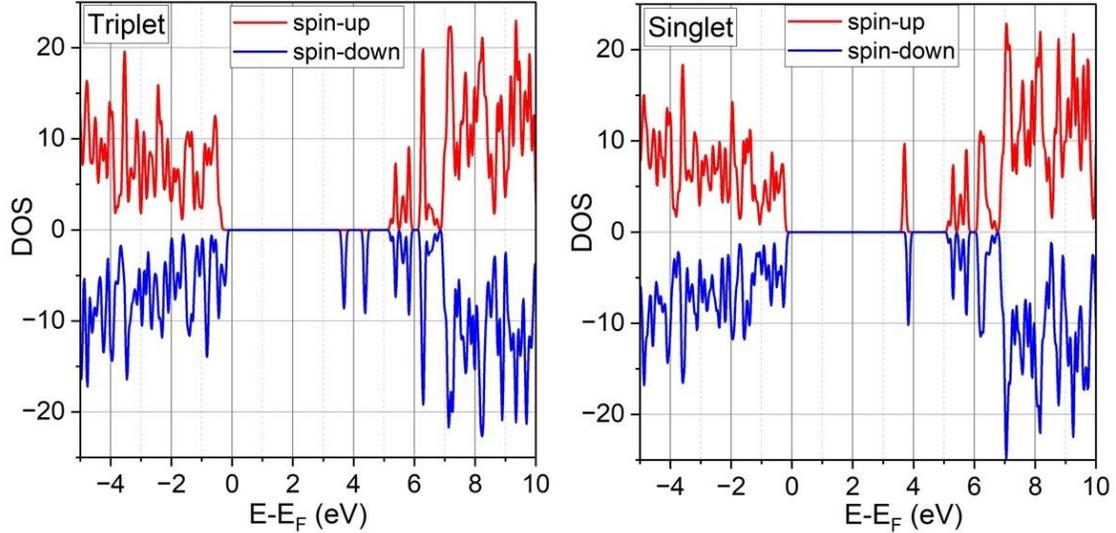

FIG. 4. Density of the IQP states calculated for the triplet and singlet state of the $V_BC_B$ defect.

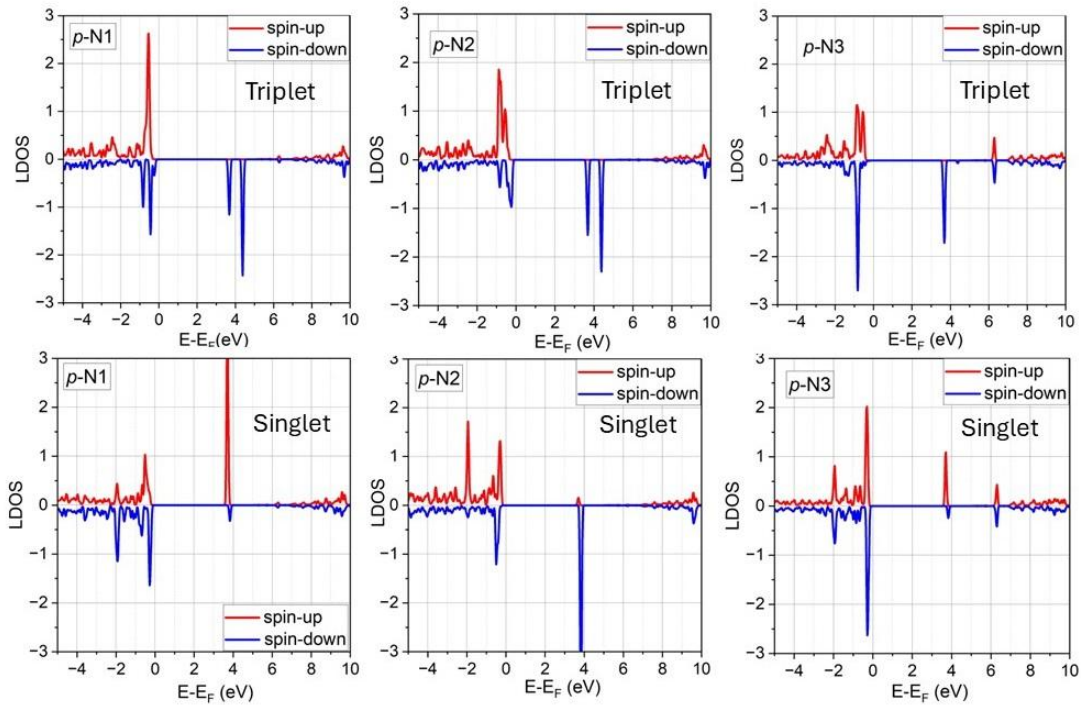

FIG. 5. Local densities of the *p*-states of the undercoordinated N atoms calculated for the triplet and singlet in the $V_BC_B$ defect.

We apply the BSE method to calculate the frequency-dependent dielectric function and oscillator strength of optical dipole excitations for the triplet state of $V_BC_B$. The results are shown in the left panel of Fig. 6. We find that a sharp excitation peak is formed in the triplet at an energy of about 3.2 eV. This transition has a oscillator strength comparable or higher than that of other defects in hBN that we studied earlier. [21] This excitation is the first step of the spin-polarization cycle. The next step is expected to be a phonon-assisted ISC to a singlet state. It cannot occur directly from the excited triplet state to the ground singlet state because of the large difference in energy (about 3 eV). However, it can proceed through some singlet excited state, as it happens in the $NV^-$ center. To obtain the singlet excited states we use the BSE method to calculate its optical excitation spectrum. The results are shown in the right panel of Fig. 6. We find the singlet to have several sharp defect-related excitation peaks. The peak at 2.46 eV corresponds to the singlet excited state that may contribute to the spin-polarization cycle.

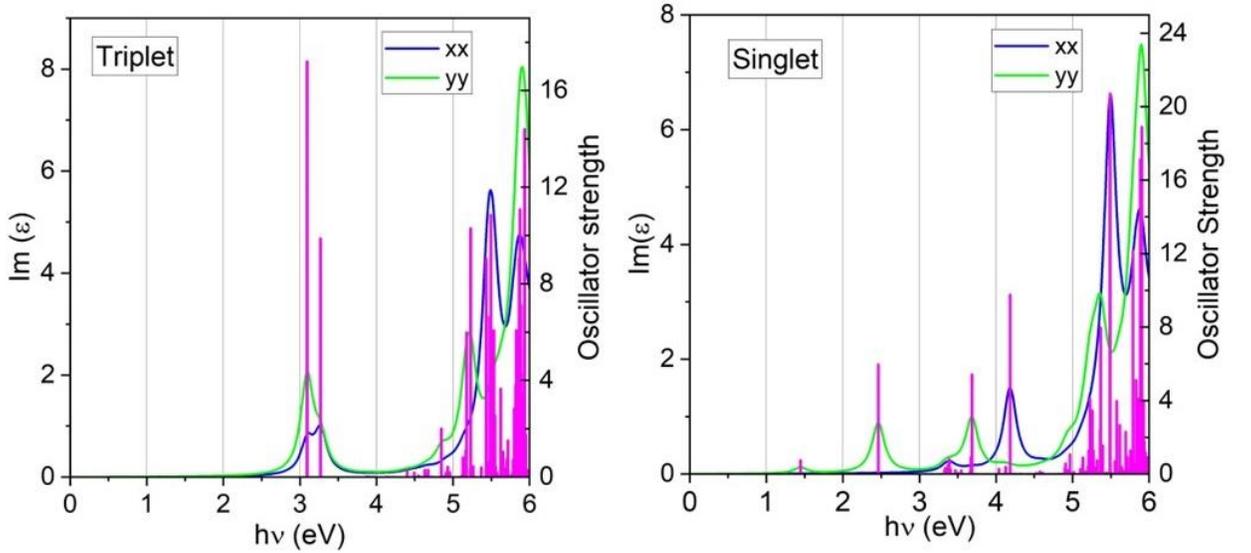

FIG. 6. The frequency-dependent dielectric function (blue and green lines) and the oscillator strength (pink bars) calculated for the triplet and singlet spin states in the $V_BC_B$ defect. The zz-polarization of Im($\varepsilon$) is not shown because its magnitude is negligible.

### C. Spin-polarization cycle proposed for the $V_BC_B$ defect

We find out from our calculations that the $V_BC_B$ defect has stable triplet and singlet states and that the total energy of the singlet ground state energy is 0.044 eV higher than that of the triplet ground state. Our BSE calculations indicate that the singlet has excited states with energy lower than those of the triplet. These conditions are favorable for the spin-qubit optical initialization via the cycle described in the Introduction. We thus construct the spin-polarization cycle for $V_BC_B$ (see Fig. 7). The first step in the loop is the optical excitation from the triplet ground state to the triplet excited state. We find two excitations relevant to the loop with energies around 3.2 eV. This excitation will require a UV laser. Next, we can expect the system to undergo the phonon-induced ISC from the triplet excited to the singlet excited state with $m_S = \pm 1$ (Via a selection rule) followed by emission to the singlet ground state and then another ISC to the triplet ground state. As seen from Fig. 7, the relative energies of the states involved in the process make the spin-polarization cycle achievable.

The proposed cycle in the $V_BC_B$ defect may be more efficient for technological applications than that in the $V_B^-$ defect. Namely, the high rate of the ISC from the triplet excited to the singlet excited states of $V_B^-$ has been found to be responsible for short triplet excited-state lifetime and its very low quantum yield. 9]. This has been supported by calculations presented Ref. in [11]. The authors found that the triplet to singlet excited-state transition has a rate several orders of magnitude larger rate than that calculated for the triplet's optical decay for the $m_s$=0 spin projection. Importantly, they traced the high rate of the ISC from the triplet excited to the singlet excited state to their near degeneracy, their similar structure, and strong spin-orbit coupling. Nevertheless, such degeneracy does not happen for the proposed $V_BC_B$ defect. We thus expect the spin qubit initialization to be more efficient in $V_BC_B$ than in $V_B^-$. Another important characteristic of spin qubits is the zero-field splitting (ZFS), which is the splitting between the triplet spin projections caused by the dipolar spin-spin interaction. We calculated the ZFS parameter D for the triplet ground state and found it to be equal to 2.706 GHz. This value is comparable to the ZFS parameters obtained for the $V_B^-$ D = 3.48 GHz, [10] and NV- defects D = 2.88 GHz, [23] indicating that the spin polarization manipulation is feasible as it is for other studied defects. We thus conclude that the proposed $V_BC_B$ defect in hBN has properties favorable for the qubit functionalities including its initialization, manipulation, and readout.

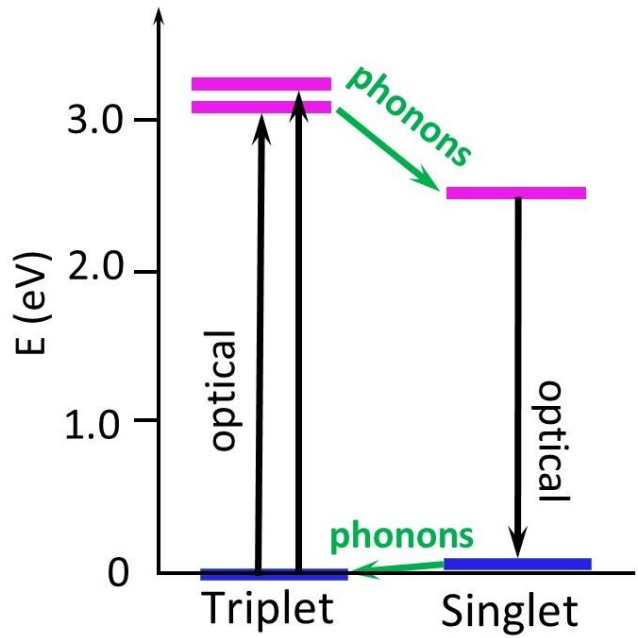

FIG. 7. The proposed energy diagram for the spin-polarization cycle in the $V_BC_B$ defect. The blue and pink bars indicate the ground and excited states, respectively.

## IV. CONCLUSIONS

By applying the existing knowledge and an educated guess, we proposed the $V_BC_B$ defect in hBN as a possible spin qubit. Our DFT calculations confirmed that the defect has stable triplet and singlet ground states, and that the latter has a total energy 0.044 eV higher than the former, which is favorable for the spin-qubit functionality. We found from our GW calculations that the spin-active electronic states associated with the defect are localized around the undercoordinated N atoms and create sharp peaks in the density of electronic states within the bandgap. The frequency-dependent dielectric function and oscillator strength have been calculated for the triplet and singlet within the BSE method. We found from these calculations that the optical excitations in both the triplet and the singlet have energies and rates favorable for the spin-polarization cycle. Using the data obtained for the ground and excited state energies; we constructed a spin-polarization-cycle diagram. This diagram, along with the result of the ZFS calculations, indicates that the $V_BC_B$ defect has promising spin qubit functionalities.

**Acknowlegment** This work was supported by the U.S. Department of Energy, Office of Science, Basic Energy Sciences, under Award # DE-SC0024487.